\documentclass[aps,prb,onecolumn,notitlepage]{revtex4-1}%

\usepackage{graphicx}
\usepackage{amsmath,amstext, amssymb}   
\usepackage{amsfonts}
\begin{document}
\title{Ultrafast Hole-Spin Dynamics in Optically Excited Bulk GaAs} 
\author{Michael Krau\ss} 
\affiliation{Department of Physics and Research Center OPTIMAS,
  University of Kaiserslautern, 67663 Kaiserslautern,
  Germany} 
\author{David J. Hilton}
\email{dhilton@uab.edu} 
\affiliation{Department of Physics, University of Alabama at Birmingham, 
Birmingham, AL, 35294-1170}
\author{Hans Christian Schneider}
\email{hcsch@physik.uni-kl.de} 
\affiliation{Department of Physics and Research Center OPTIMAS,
  University of Kaiserslautern, 67663 Kaiserslautern,
  Germany} 
\date{\today}


\begin{abstract}
  We present experimental and theoretical results on hole-spin
  dynamics in bulk GaAs after ultrafast optical excitation. The
  experimental differential transmission are compared with a dynamical
  calculation of the momentum-resolved hole distributions, which
  includes the carrier-carrier, carrier-phonon and carrier-impurity
  interaction at the level of Boltzmann scattering integrals. We
  obtain good agreement with the experimentally determined hole-spin
  relaxation times, but point out that depending on how the
  spin-polarization dynamics is extracted, deviations from an
  exponential decay at short times occur. We also study theoretically
  the behavior of the spin-relaxation for heavily p-doped GaAs at low
  temperatures.
\end{abstract}

\maketitle

Research on spin dynamics and spintronics in semiconductors and metals
covers an immense variety of applications in
information storage and manipulation in carrier spins in solid state
systems.~\cite{awschalom-book,wolf:science,zutic:review} Both storage
and manipulation of spins are limited by spin
relaxation, so that an understanding of spin relaxation processes is
important.~\cite{sham:review93,averkiev:review02} For ultrafast spin
dependent dynamics on picosecond timescales, simplified
relaxation-time approximations are no longer justified and a fully
\emph{microscopic} understanding of the complex spin dynamics is
needed.~\cite{glazov-jetplett02,wu:prb03,hcs-prb06:spin-relax-surface}
While electron spin-dynamics continue to be extensively investigated
in semiconductors due to their extremely long spin-relaxation
times,~\cite{kikkawa:prl98} hole-spin dynamics are intriguing for
different reasons. In III-V semiconductors, hole dynamics
is inherently different from that of electrons because of the strong
spin-orbit coupling of the hole states. Thus, hole-spin relaxation and
momentum (energy) relaxation occur on a comparable ultrashort
timescale. 

In this paper, we connect differential transmission measurements of
the hole-spin dynamics~\cite{hilton-tang} with a microscopic
calculation that includes the anisotropic band-structure as well as
carrier-carrier and carrier-phonon/impurity scattering mechanisms at
the level of Boltzmann scattering integrals. Although our approach
does not take into account all the intricacies of the dynamical
multi-band screening (including phonon-plasmon
coupling)~\cite{Collet:1993p794,Woerner:1995p20} and interband
polarizations, it captures essential aspects of the spin dynamics of
holes.~\cite{krauss-prl08} We show that this calculation compares well
with measured hole-spin relaxation times, but that deviations from an
exponential spin decay occur for short times depending on how the
relaxation time is extracted. Further, we present results for the
hole-spin relaxation at high p-doping, which underlies a recent
treatment of GaAs-based ferromagnetic
semiconductors.~\cite{Cywinski-prb07:sp-d-model}

Recent theoretical treatments of hole spin (and charge)
dynamics in bulk GaAs have focused on the calculation of
energy-dependent spin \emph{relaxation-rates} due to phonon/impurity
scattering,~\cite{schilf-prb05} the investigation of coherent hole
(spin) dynamics in GaAs,~\cite{Dargys:2004p403,winkler:prl06} and the
influence of Dyakonov-Perel type precession effects due to
spin-dependent splittings between hole bands in GaAs quantum
wells.~\cite{wu:prb06:holespin} Very recently, additional
contributions due to ``small'' band structure effects and the
spin-orbit contribution to the hole-phonon interaction for holes in
bulk GaAs have also been analyzed.~\cite{wu:holes-bulkGaAs}

From an experimental point of view, hole-spin polarizations can be
created efficiently using optical orientation
techniques.~\cite{opt-orient} Because of the ultrashort
lifetimes~\cite{Ganikhanov:1999p12} involved, it is
difficult to \emph{unambiguously} study hole dynamics in degenerate
pump-probe experiments (i.e., experiments with the same pump and probe
wavelengths) due to the competing presence of the coherent artifact on
the same timescale\cite{Luo:2009p6827} and, to a lesser degree, the
electron spin dynamics.\cite{kikkawa:prl98} Hole spin study,
therefore, requires measurement of the spin-dependent carrier
dynamics on ultrashort timescales using a time-delayed, mid-infrared
probe beam (i.e., a non\-degenerate pump-probe experiment), which was
realized only much later~\cite{hilton-tang} than similar experiments
on electrons.~\cite{kikkawa:prl98}

We focus here on the dynamics of the spin-polarized heavy hole
subband.  We have performed non\-degenerate, polarization-resolved
pump-probe spectroscopy at 300\,K to measure the hole spin relaxation
lifetime in intrinsic GaAs as a function of input pump fluence,
$\mathcal{F}$.  In this setup, 1\,W of average power from an 80\,fs
titanium:sapphire laser (Spectra Physics Tsunami) operating at an
80\,MHz repetition rate is used to pump a custom-designed optical
parametric oscillator based on periodically-poled lithium niobate, as
described in further detail in ref.~\onlinecite{Burr:1997p7165}.  A
portion of the titanium:sapphire beam ($\le 300$\,mW) is circularly
polarized using a quarter wave plate and is used to photo\-excite
spin-polarized electron-hole pairs in the GaAs sample. The pump beam
induces a circular birefringence and results in a time-dependent
rotation of the transmitted probe polarization.  To isolate the
dynamics of the hole spin state from the electron spin, we probe the
occupancy of the heavy hole and/or light hole subbands using a
synchronized mid-infrared idler pulse ($\lambda$ = 3200 nm), which
probes the induced transitions from the split off subband into the
heavy hole subband.  This wavelength directly probes states from the
lower split-off hole subband to the heavy hole subband at the same
quasimomentum ($k\approx 0.3~\mathrm{nm}^{-1}$) as the initial
photoexcitation at 800\,nm.

\begin{figure}[tb!]
\includegraphics[angle=-90, width=0.4\textwidth]{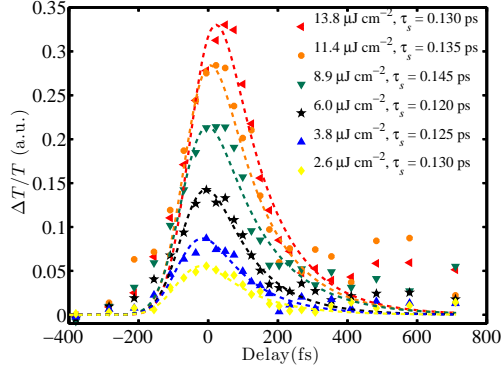}
\caption{Fluence dependence of the time-resolved differential
  transmission, $\Delta T/T(t)$, at 300\,K.
\label{fig:fluence}}
\end{figure}

Figure~\ref{fig:fluence} shows the measured field rotation, $\Delta
T/T\bigl(t\bigr)$, of the mid-IR probe from $\mathcal{F}$ = 2.6 to
13.8 $\mu$J cm$^{-2}$.  The input probe beam is linearly polarized
($\hat{ x}$), which is an equal admixture of both the co-circularly
polarized ($\hat{\sigma}_{+}$) and counter circularly
polarized($\hat{\sigma}_{-}$).  A MgF$_2$ linear polarizer is rotated
$90^{\circ}$ with respect to the input polarization (i.e., along
$\hat{y}$) and is used to measure the field rotation of the
transmitted probe, which depends on the time-dynamics of the the heavy
hole occupancy state near the quasi\-momentum, $k$, of
photo\-generation. Using the measured absorption coefficient of
GaAs,~\cite{blakemore} the resulting carrier concentrations range from
$5\times10^{16}\,\mathrm{cm}^{-3}$ ($\mathcal{F} =
2.6~\mu\mathrm{J\,cm}^{-2}$) to $2\times10^{17}\,\mathrm{cm}^{-3}$
($\mathcal{F} = 13.8~\mu\mathrm{J\,cm}^{-2}$). We fit the measured
field rotation to a single exponential model and account for the
finite pump and probe pulse widths by convolving the model function
with a gaussian with a width of 120\,fs (see
ref.~\onlinecite{hilton-tang} for more details of the fitting model).
We find no significant trend in the extracted lifetimes, which range
from 120\,fs to 145\,fs, within the experimental error ($\pm 20\%$) of this
experiment.

The calculation of the hole-spin dynamics follow the procedure outlined in
ref.~\onlinecite{krauss-prl08}. The electron and hole states around
the fundamental bandgap needed as input for the dynamical calculation
are determined from an 8-band Kane model $\mathcal{H}(\vec k)$ with six
hole and two electron bands containing terms up to second order in
$k$.~\cite{winkler:book}  By diagonalizing $\mathcal{H}(\vec{k})$ we
obtain the ``intelligent basis'' (in the sense of
ref.~\onlinecite{Fabian:review}) for the hole states: $\lvert\nu,\vec
{k}\rangle$ and their energy eigenvalues
$\varepsilon_{\nu,\vec{k}}$. The label~$\nu = (b,p)$ includes the band
index $b=$ E, HH, LH, SOH, (for electrons, heavy holes, light holes,
and split-off holes) and the pseudospin $p=1,2$. The pseudospin can be
introduced because the quasiparticle dispersions of all four types of
carriers are (nearly) doubly degenerate. 

The dynamical equations governing the time evolution of the incoherent
carrier distributions, $n_{\nu, \vec k}$, under the influence of
optical fields and scattering are:~\cite{collet:93,krauss-prl08}
\begin{equation}
\frac{\partial}{\partial t}
n_{\nu,\vec{k}}=\Gamma_{\nu,\vec{k}}
^{\mathrm{in}}(1-n_{\nu,\vec{k}})-\Gamma_{\nu,\vec{k}}^{\mathrm{out}}
n_{\nu,\vec{k}}
\end{equation}
 (plus a contribution from the optical excitation,
which neglects the hole band coherences between hole states of the
``intelligent basis,'' or, equivalently, only describes contributions
to the hole spin relaxation of the Elliott type~\cite{Fabian:review}
that arise from the $\vec{k}$ dependent mixing of different spins in
the states $\lvert \nu, \vec{k}\rangle$.  Contributions to the spin
relaxation via coherent spin precession,i.e., Dyakonov-Perel-type spin
dynamics,~\cite{wu:holes-bulkGaAs} are neglected.

The dynamical in-scattering rate consists of the carrier-carrier
interaction contribution (i.e.,  eq.~(2) in
Ref.~\onlinecite{krauss-prl08}) and the carrier-phonon interaction
\begin{equation}
\Gamma^{\mathrm{in}}_{\nu,\vec k}|_{\mathrm{c-p}}
=\frac{2\pi}{\hbar} \sum_{\nu_{1},\vec{q},\lambda}
|M_{q,\lambda}|^2|\langle\nu_1,\vec{k}+\vec{q}|\nu,\vec{k}\rangle|^2 
n_{\nu_1,\vec{k}+\vec{q}}
 \left[ (1+N_{q,\lambda})
\delta(\Delta E_{-}) + N_{q,\lambda}
\delta(\Delta E_{+})\right] \ .
\label{gamma-in-cp} 
\end{equation}   
where $\Delta E_{\pm} = \varepsilon_{\nu, k}
-\varepsilon_{\nu_1,|\vec{k}+\vec{q}|} \pm \hbar \omega_{q,\lambda}$.
There is also a contribution similar to eq.~\eqref{gamma-in-cp} due to
carrier-impurity scattering.~\cite{wu:prb03} Here, $N_{q,\lambda}$,
$M_{q,\lambda}$, and $\omega_{q,\lambda}$ are the phonon occupation
numbers, carrier-phonon coupling matrix elements, and phonon
dispersions for LO and LA phonons, respectively.~\cite{yu-cardona} The
out-scattering rates $\Gamma^{\mathrm{out}}$ are obtained from
$\Gamma^{\mathrm{in}}$ by exchanging $(1-n)$ with~$n$ and~$(1+N)$
with~$N$. Equation~\eqref{gamma-in-cp} and its counterpart for Coulomb
scattering describes two-particle scattering processes connecting
states $\lvert\nu,\vec
{k}\rangle\rightarrow|\nu_{1},\vec{k}+\vec{q}\rangle$ (and
$\lvert\nu_{2},\vec
{k}_{1}+\vec{q}\rangle\rightarrow|\nu_{3},\vec{k}_{1}\rangle$) with
different average spin.  The pronounced anisotropy of the
single-particle states, which is due to the spin-orbit interaction, is
included in our calculation in the overlaps and the carrier energies,
$\epsilon_{\nu, \vec k}$, which enter the energy-conserving delta
functions in the scattering rates.  The anisotropies and
high-dimensional scattering integrals result in a challenging
numerical problem. We incorporate the effects of the anisotropy by
expanding $n_{\nu,\vec{k}}(t)$ into spherical harmonics
$Y_{\ell,m}(\hat{k})$ up to order $\ell=4$ and retain only the
expansion coefficients with radial or cubic symmetry, as these are the
dominant symmetries of the
Hamiltonian~$\mathcal{H}$.~\cite{winkler:book} This procedure,
together with the use of static screening, $v^s(q) =
e^2/[\epsilon_0\epsilon_{bg}(q^2 + \kappa^2)$], reduces the numerical
complexity. Additional simplifications are achieved by including only
the heavy-hole bands in the dynamical calculation as the number of
optically excited light holes is much smaller,~\cite{krauss-prl08} and
by replacing the non\-equilibrium electron distributions by
equilibrium distributions with the lattice temperature. We then
calculate the spin polarization of the HH band. In
ref.~\onlinecite{krauss-prl08}, it was shown that the spin
polarization is not identical to the DT signal that is measured, but
the deviations are small enough for the accuracy of the present
comparison.

We include the carrier excitation by an ultrashort optical pulse using
a $\hat\sigma_{+}$ polarized plane wave traveling in the (001)
direction ($\hat{z}$) to generate an initial condition for the carrier
distributions
\begin{equation}
n_{\nu,\vec{k}}(t=0)=\sum_{\mu}|\vec{d}_{\mu\nu}(\vec{k})\cdot\vec{E}|^{2}
g(\hbar\omega-\varepsilon_{\mu,\vec{k}}-\varepsilon_{\nu,\vec{k}})\ .
\end{equation}
Here, $\vec{d}_{\mu\nu}(\vec{k})=e\langle
\mu,\vec{k}|\vec{r}|\nu,\vec{k}\rangle$ are the dipole-matrix
elements, $\vec{E}$ is the laser field, and $\hbar\omega$ is the
photon energy of the exciting field. A Gaussian broadening function
$g\bigl(\varepsilon\bigr)$ peaked at $\varepsilon =
\bigl(\hbar\omega-\varepsilon_{\mu,\vec{k}}-\varepsilon
_{\nu,\vec{k}}\bigr)$ accounts for the spectral width of the pulse ($
15$\,meV).

\begin{figure}[t]
\includegraphics[width=0.4\textwidth]{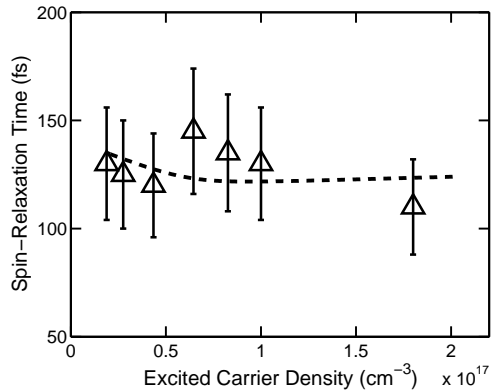}
\caption{Theoretical (curve) and experimental values (symbols) for
  fluence dependence of hole-spin relaxation in intrinsic GaAs at room
  temperature.}
\label{fludep}
\end{figure}

To compare with experimental results in Fig.~\ref{fludep}, we convert
the experimental laser fluence into excited carrier densities, and
extract the relaxation time from an exponential fit to the spin
polarization at the probe-laser wavelength. The experimental results
are in good agreement with our calculations and a recent theoretical
study.~\cite{wu:holes-bulkGaAs}. Fig.~\ref{fludep} shows that hole
spin-relaxation is rather insensitive to excited carrier densities on
the order of $10^{17}$\,cm$^{-3}$. The calculation also predicts a
very weak temperature dependence with a hole spin relaxation time of
250\,fs for an excited density of $10^{17}$\,cm$^{-3}$ at 4\,K, which
is somewhat shorter than the result predicted by the Elliott-Yafet spin
\emph{relaxation-time}.~\cite{schilf-prb05}

\begin{figure}[t!]
\includegraphics[width=0.4\textwidth]{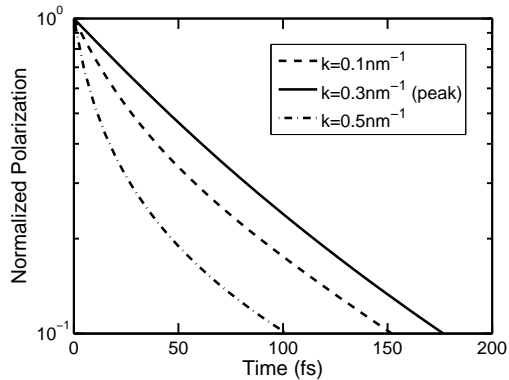}
\caption{Computed hole-spin polarization dynamics at different hole
  momenta for fixed excitation at $k = 0.3\,\mathrm{nm}^{-1}$ in
  intrinsic bulk GaAs. The excited carrier density is
  $n=10^{17}$\,cm$^{-3}$.}
\label{wavdep}
\end{figure}

In Fig.~\ref{wavdep}, we simulate the momentum dependence of the spin
polarization dynamics. We model the excitation by an 800\,nm pump
pulse that excites a carrier density of
$10^{17}\,\mathrm{cm}^{-3}$. This leads to anisotropic initial hole
distributions that are peaked at momentum $k = 0.3\,\mathrm{nm}^{-1}$
(after integration over the angular variables). Fig.~\ref{wavdep}
shows different spin polarizations calculated using the angle-averaged
distributions~$n(k,t)$ for different modulus of the hole momentum
$k$. The decay of the spin polarization near the peak position of the
initial distribution is nearly exponential, as can be seen from the
linear curve shape on the logarithmic scale. The evolution of the spin
polarization for larger and smaller momenta (i.e, away from the
momenta of the initially photoexcited holes) is significantly faster
and non-exponential for the initial $\sim 100$\,fs. After the initial
non-exponential dynamics the spin-relaxation becomes exponential again
with the same time constant. The spin relaxation-time therefore
depends on the way it is extracted from the calculation, as opposed to
the case of electron spin-relaxation in p-doped GaAs where the spin
relaxation time can be quite rigorously defined.~\cite{SPIE} The
dashed lines in Fig.~\ref{wavdep} correspond to experimental
situations where pump and probe pulses are detuned, as was
investigated in ref.~\onlinecite{hilton-tang} (see Fig.~5 therein),
where no change in spin relaxation times for different detuning
between pump and probe pulses was found. The apparent discrepancy
likely results from the analysis of the experimental data via a
convolution of Gaussians, which are used to describe both the pulse
shapes and the dynamics of the material response. The analysis of the
pump-probe experiment therefore presupposed an exponential decay of
the polarization, which turns out to be only somewhat longer than the
experimental time resolution of about 100\,fs~\cite{hilton-tang}. For
degenerate pump and probe, this analysis yields the exponential decay
of the spin polarization in agreement with the calculated result in
Fig.~\ref{wavdep}. For non-degenerate situations the non-exponential
features in Fig.~\ref{wavdep} on timescales below 100\,fs are not
resolved. Their resolution would require a time resolution well below
100\,fs.

\begin{figure}[t!]
\includegraphics[width=0.4\textwidth]{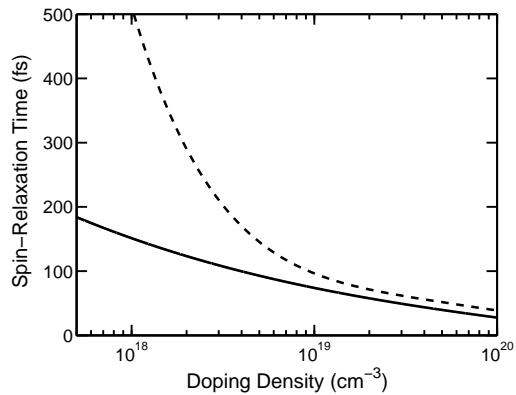}
\caption{Computed p-doping dependence of hole-spin relaxation in GaAs
  at low temperatures (solid line). The dashed line results if only
  scattering by from ionized donors is included.}
\label{dopdep}
\end{figure}

Finally, we examine the hole-spin relaxation dynamics in heavily
p-doped GaAs at low temperatures (5\,K) theoretically. By choosing
these parameters, we estimate the hole-spin relaxation time for GaMnAs
in the ferromagnetic phase at low temperatures, which has similar
carrier doping levels.  We model the optical excitation of
nonequilibrium carriers in p-doped GaAs by a pump laser with a
wavelength of 400\,nm. Choosing this wavelength ensures that the
optical transitions are not blocked by holes even for very high doping
densities. We assume initial carrier distributions of the form $n^0_k
= n^{\mathrm{dop}}_k + n^{\mathrm{exc}}_k$, which contain both the
influence of the doping and the optical excitation. All itinerant
holes introduced by the p-doping are assumed to be thermalized at the
lattice temperature; they are modeled by unpolarized Fermi-Dirac
electron distributions $n^{\mathrm{dop}} = f(\epsilon^{\mathrm{e}}_k)$
at lattice temperature with carrier density equal to the density of
dopants. For the optical excitation we assume again an ultrashort
pulse with a width of $15$\,meV. Fig.~4 shows that the hole-spin
relaxation drops to approximately 10\,fs for doping densities
typical for GaAs based magnetic semiconductors. At these high doping
densities, the rapid momentum relaxation due to scattering from the
ionized donors provides the main relaxation mechanism. The assumption
of a hole spin relaxation time of $\sim$10\,fs in
ref.~\onlinecite{Cywinski-prb07:sp-d-model} makes it possible to
neglect a spin-bottleneck effect for ferromagnetic GaMnAs. Our
results, based on a dynamical calculation, support this assumption.

\newpage

In conclusion we examined ultrafast hole-spin relaxation using
differential transmission measurements and a microscopic theoretical
approach for different excitation conditions. Theory and experiment
show that the hole-spin relaxation time is approximately 100 fs and
independent of moderate changes of the pump fluence at room
temperature. The concept of spin-relaxation time is not, in general,
applicable for non-degenerate pump-probe schemes, but the deviations
from the exponential decay of the spin-polarization are confined to
time scales of less than 100\,fs, and can therefore only be resolved
by a time resolution well below 100\,fs. These results also
demonstrate the limited validity of a single spin
\emph{relaxation-time} for holes. Strong p-type doping shortens
relaxation times significantly due to the scattering from ionized
donors. The calculated spin relaxation times support the assumption of
very fast hole-spin relaxation in ferromagnetic GaMnAs at low
temperatures.

We acknowledge support by the DFG through GRK 792 and a grant for CPU
time from the the NIC J\"{u}lich. We are grateful to C.~L.~Tang and
M.~W.~Wu for helpful discussions.

\bibliography{dephasing2,hcs_pubs2,spin3,djhadded,new}

\end{document}